\documentclass{aastex}
\usepackage{emulateapj5}
\usepackage{onecolfloat5}
\usepackage{apjfonts}

\def\plotone#1{\centering \leavevmode
\epsfxsize= 1.0\columnwidth \epsfbox{#1}}

\def\gsim{\;\rlap{\lower 2.5pt
 \hbox{$\sim$}}\raise 1.5pt\hbox{$>$}\;}
\def\lsim{\;\rlap{\lower 2.5pt
   \hbox{$\sim$}}\raise 1.5pt\hbox{$<$}\;}

\newcommand{\be}{\begin{equation}}
\newcommand{\beq}{\begin{equation}}
\newcommand{\ba}{\begin{eqnarray}}
\newcommand{\ee}{\end{equation}}
\newcommand{\eeq}{\end{equation}}
\newcommand{\ea}{\end{eqnarray}}

\usepackage{epsfig}

\begin{document}
\twocolumn[%%% Begin front material
%\submitted{Submitted to ApJ}
\title{Reconstructing the Cosmic Evolution of Quasars from the \\ Age Distribution of Local Early--Type Galaxies}

\author{Zolt\'an Haiman}
\affil{Department of Astronomy, Columbia University, 550 West 120th Street, New York, NY 10027, USA; zoltan@astro.columbia.edu}

\author{Raul Jimenez \&  Mariangela Bernardi}
\affil{Department of Physics and Astronomy, University of Pennsylvania, 209 South 33rd Street, Philadelphia, PA 19104, USA; (raulj,bernardm)@physics.upenn.edu}

\begin{abstract}
  We use the spectra of $\approx$22,000 nearby early--type galaxies
  from the Sloan Digital Sky Survey (SDSS) to determine the age
  distribution of these galaxies as a function of their velocity
  dispersion $\sigma_v$ in the range $100~{\rm km~s^{-1}} \lsim
  \sigma_v \lsim 280~{\rm km~s^{-1}}$.  We then combine the inferred
  age--distributions with the local abundance of spheroids, including
  early--type galaxies and late--type bulges, to predict the evolution
  of the quasar luminosity function (LF) in the redshift range
  $0<z\lsim 6$.  We make the following simple assumptions: (i) the
  formation of stars in each galaxy, at the epoch identified with the
  mean mass--weighted stellar age, is accompanied by the prompt
  assembly of the nuclear supermassive black hole (SMBH); (ii) the
  mass of the SMBH obeys the $M_{\rm bh}-\sigma_v$ correlation
  observed in nearby galaxies; (iii) the SMBH radiates at a fraction
  $f_{\rm Edd}$ of the Eddington limit for a fixed duration $t_{\rm
  Q}$, and is identified as a luminous quasar during this epoch, (iv)
  the intrinsic dispersions in the Eddington ratio and the $M_{\rm
  bh}-\sigma_v$ relation produce a combined scatter of $\Delta \log
  L_{\rm Q}$ around the mean logarithmic quasar luminosity $\langle
  \log L_{\rm Q} \rangle$ at fixed $\sigma_v$.  These assumptions
  require that the SMBH remnants of quasars with $L_{\rm bol}\lsim
  10^{12.5} f_{\rm Edd}{\rm L_\odot}$ reside predominantly in bulges
  of late type galaxies.  We find that evolution of the observed
  quasar LF can be fit over the entire redshift range in this simple
  model, $0<z\lsim 6$ with the choices of $0.6\lsim \Delta \log L_{\rm
  Q}\lsim 0.9$, $6\times 10^7 {\rm yr} \lsim t_{\rm Q}\lsim 8\times
  10^7$ yr, and $0.3\lsim \langle f_{\rm Edd}\rangle \lsim 0.5$.
  We find no evidence that any of the model parameters evolves with
  redshift, supporting the strong connection between the formation of
  stars and nuclear SMBHs in spheroids.
\end{abstract}
%\vspace{3\baselineskip}
]%%% End front material

\section{Introduction}
\label{sec:introduction}

The discovery of tight correlations between the masses of supermassive
black holes (SMBHs) at the centers of galaxies and the global
properties of the spheroid component of the galaxies themselves (see,
e.g., Magorrian et al. 1998; Ferrarese \& Merritt 2000; Gebhardt et
al. 2000; Graham et al. 2001) have suggested a strong link between the
formation of SMBHs and their host spheroids.  Several groups have
noted that galaxy formation and SMBH growth should be linked, and many
modeled the joint cosmological evolution of quasars and galaxies (see,
e.g., Monaco et al. 2000; Kauffmann \& Haehnelt 2001; Haiman \& Menou
2001; Granato et al. 2001; Ciotti \& van Albada 2001, Cavaliere \&
Vittorini 2002; Lapi et al. 2006, and references therein).

A particularly revealing characteristic of this cosmic evolution is
the so--called ``downsizing''. Observations have long suggested an
inverse correlation between the ages of elliptical galaxies and their
size, with more massive ellipticals forming at earlier epochs (e.g.,
Heavens et al. 2004 and references therein).  Interestingly, the
characteristic quasar luminosity similarly declines over time from
$z\approx 2$ to $z=0$ (Boyle et al. 2000).  At first sight, the
decline of both the characteristic galaxy velocity dispersion, and of
quasar luminosity, is surprising in ``bottom--up'' hierarchical
structure formation models.  This anti--hierarchical behavior, can
nevertheless be naturally accommodated in these structure formation
models, invoking the energy feedback from the AGN on the host galaxy
(e.g. Granato et al. 2004) and large-scale environmental effects
(e.g. Blanton et al. 1999).  The similarity between the ``downsizing''
observed for elliptical galaxies and for quasars has been noted in
previous work (e.g. Franceschini et al. 1999; Haiman, Ciotti \&
Ostriker 2004), and, as these works pointed out, it suggests that the
formation of SMBHs and spheroids was linked by the same physical
mechanism at all cosmic epochs.

The simplest hypothesis to interpret the above trends is that star
formation in spheroids and BH fueling are proportional to one another
at all times (Haiman, Ciotti \& Ostriker 2004).  With the additional
assumptions that BH accretion luminosity stays at some fraction of the
Eddington limit during luminous quasar phases, and that these
phases have a characteristic duty cycle, one can predict the evolution
of quasars from the evolution of early--type galaxies (and vice versa).
Unfortunately, a direct observational test of the above hypotheses is
not possible, as the cosmic evolution of the population of early--type
galaxies has not been determined from observations out to high $z$.

Here we follow an alternative ``archaeological'' approach, and we use
age--distributions we infer for nearby early--type galaxies, as a
function of their velocity dispersion, to reconstruct in detail the
evolution of quasars, as a function of their luminosity.  This
approach has been attempted previously by Cattaneo \& Bernardi (2003),
who have used the mean age of early--type galaxies {\it vs} their
velocity dispersion, together with assumptions about the geometry of
obscuring tori, to infer the evolution of their abundance, and to
predict the evolution of the quasar LF (see also Granato et al. 2001
for the inverse approach). In this paper, we improve on these previous
studies in several ways: (i) we infer the age distribution, rather
than only the mean age, as a function of $\sigma_v$, for early type
galaxies, (ii) we include the population of late--type bulges in our
analysis, which indeed dominate the spheroid population at low
velocity dispersion, (iii) we perform a quantitative statistical
comparison between the predicted and observed quasar LFs, and (iii)
the quasar lifetime, the mean Eddington ratio of SMBHs, and its
scatter, are taken as three fitting parameters that are fit
simultaneously.  We find that the shape of the quasar LF we predict is
in rough agreement with observations, without introducing either a
luminosity--dependent obscuration or a redshift--evolution of either
of the three model parameters.  Our results require that the quasar
luminosity $L_{\rm Q}$ has a non--negligible scatter (of $\gsim 0.5$
dex) at a fixed value $\sigma_v$ of the host galaxy.

The rest of this paper is organized as follows.
In \S~\ref{sec:ages}, we discuss the sample of early type galaxies
we used in our analysis, and infer the age distribution of these galaxies.
In \S~\ref{sec:quasars}, we discuss our modeling that converts the
inferred age--distributions to the evolution of the quasar luminosity function
(LF).
In \S~\ref{sec:stat}, we discuss our method of statistically comparing
the predicted and observed quasar LFs.
In \S~\ref{sec:results}, we present and discuss our main results, and
in \S~\ref{sec:conclude}, we briefly summarize our conclusions.
Throughout this paper, we adopt the background cosmological
parameters $\Omega_m=0.3$, $\Omega_{\Lambda}=0.7$, and $H_0=70~{\rm
km~s^{-1}~Mpc^{-1}}$, consistent with the values measured recently by
the {\it WMAP} experiment (Spergel et al. 2006).

%\maketitle

\section{The Age Distribution of Early--Type Galaxies}
\label{sec:ages}

We have used the sample of early--type galaxies obtained by Bernardi
et al. (2006a). The sample, extracted from the Sloan Digital Sky Survey
(York et al. 2000), contains over $40,000$ early-type galaxies
selected for having apparent magnitude $14.5 \le g \le 17.75 $ with
spectroscopic parameter $eclass < 0$, which gives a PCA component
corresponding to no emission lines, and $fracDev_r > 0.8$, which is a
seeing--corrected indicator of morphology. Only those objects were
included that have measured velocity dispersions, which translates in
the spectra having $S/N > 10$ in the region $4200-5800$ \AA. The
sample extends over a redshift range $0.013 < z < 0.25$, which
corresponds to a maximum look--back time of 3 Gyr.

We computed the ages of the galaxies in three different ways. First,
for each individual galaxy, we fitted a single stellar population
model to the whole observed spectrum (continuum+lines) with two
parameters: age and metallicity. Second, we used the MOPED algorithm
(Heavens et al. 2000) to determine the star formation history of the
galaxies. Third, we used the published ages by Bernardi et al. (2006a)
that use stacked spectra of galaxies with similar properties
to measure the absorption feature produced by the Balmer absorption lines
(e.g. $H\gamma_{\rm F}$). Ages are then computed from the lines assuming
$\alpha-$enhancement with respect to the solar abundance stellar
population models. The first and third method return a
luminosity--weighted age, while the MOPED technique, because it
reconstructs the star formation history, returns the age of the
different stellar components, which can be used to computed the mass-
weighted age for the galaxy as a whole.  The three methods are
compared in detail in a companion paper by Jimenez et al. (2006).  

\begin{figure}[t]\plotone{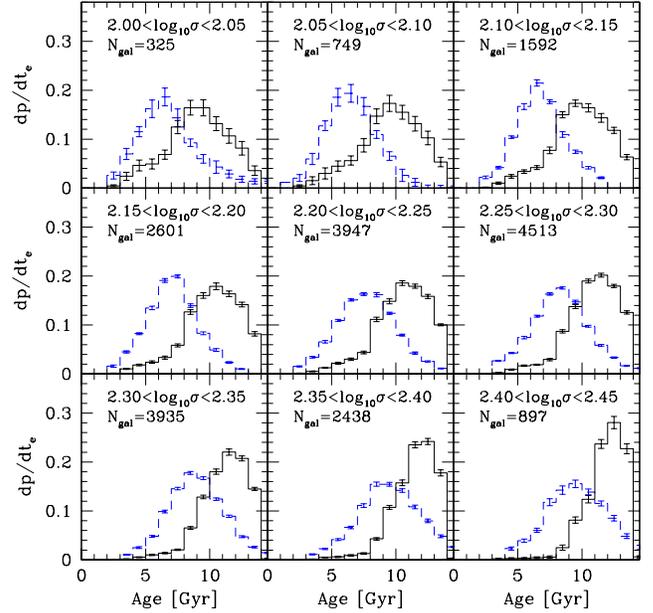}\caption{The distribution
    of stellar ages of early--type galaxies, as a function of their
    velocity dispersion $\sigma_v$ in the range $100~{\rm km~s^{-1}}
    \lsim \sigma_v \lsim 280~{\rm km~s^{-1}}$.  The spectra of $\sim
    22,000$ galaxies, all from $z<0.25$, were used from SDSS DR4. Each
    $\sigma_v$--bin contains between $\sim$300 and $\sim 4500$
    galaxies, as labeled.  The solid (black) histograms show the
    mass--weighted distributions inferred from the full spectra
    (continuum+lines) that we utilized in our analysis. Note that
    galaxies with larger $\sigma_v$ are not only older, they have a
    narrower age--distribution.  The dashed (blue) histograms show,
    for comparison, the age--distributions derived from absorption
    line indices. This method generally gives shorter ages (see text
    for discussion).  }\label{fig:agevel}\end{figure}

There are about $22,000$ galaxies in common between the MOPED sample
(Panter et al. 2006) and the Bernardi et al. (2006a) sample.  In
Fig.~\ref{fig:agevel}, we show the age distribution for several
velocity dispersions that we obtained for these galaxies.  The solid
(black) histograms show the mass--weighted age distributions from
MOPED.  The ages correspond to those extrapolated to $z=0$; note that
we impose a prior that the age of any galaxy, when extrapolated to
$z=0$, cannot exceed 13.7 Gyr (Spergel et al. 2006).  The
distributions are broad for all values of $\sigma_v$, but are narrower
for higher $\sigma_v$ and centered on older ages. The uncertainties
due to age--measurement errors are shown by the error bars in the
figure, and will be discussed in \S~\ref{sec:stat} below.

For comparison, the blue (dashed) histograms show the
age--distributions obtained from absorption line indices.  As the
figure shows, the mean ages are lower, and the shapes of the
distributions are different, from the MOPED results.  
These differences are partly accounted for by the mass vs. luminosity
weighting, and by the fact that the MOPED ages do not include a
correction for $\alpha-$enhancement. Full spectral models for the
continuum+line emission, including $\alpha-$enhancement, are not
available, but we expect that such models would somewhat reduce the
inferred ages. There appear to be additional differences between the
two age--measurement techniques (see Jimenez et al. 2006 for further
discussion). We chose to adopt the MOPED
age--distributions in our analysis below, because they utilize more
information from the spectrum, and because they can be used to compute
mass--weighted ages.  We will discuss in \S~\ref{sec:results} below
how our results would change if the ages inferred from the
absorption--line indices were used instead.

\section{Reconstructing the Quasar Luminosity Function}
\label{sec:quasars}

We use the age--distribution of the early--type galaxies, inferred
in \S~\ref{sec:ages} above, to reconstruct the evolution of the quasar
luminosity function, using the following simple assumptions.

\begin{figure}[t]
\plotone{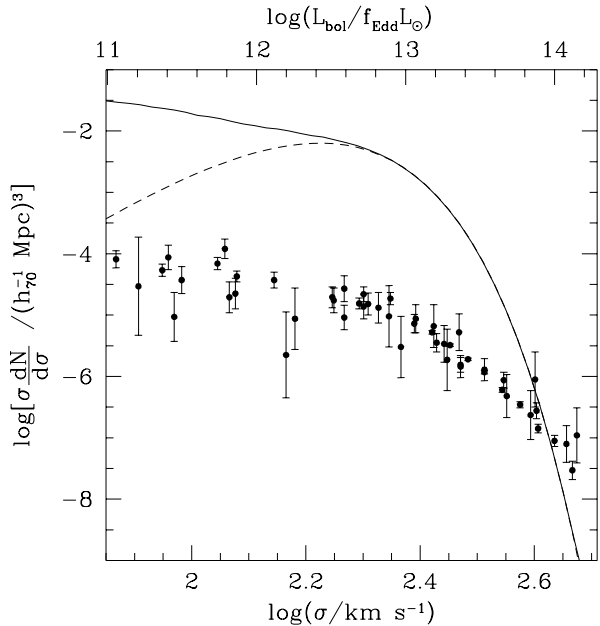}
\caption{The local velocity function spheroids, adapted from Sheth et
al. (2003), showing the contribution of early type galaxies (dashed
curve), and the whole spheroid population, including the bulges of
late type galaxies (solid curve).  For comparison, the data--points
show the quasar luminosity function at redshift $z=2.18$ (adapted from
Hopkins et al. 2006).  The upper horizontal axis labels show the mean
quasar luminosity corresponding to $\sigma_v$ from
equation~(\ref{eq:Lsigma}), which assumes that the BHs follow the
$M_{\rm bh}-\sigma_v$ relation, and shine at a fraction $f_{\rm Edd}$
of the Eddington luminosity.}
\label{fig:dndsig}
\end{figure}

We start with the velocity function of early--type galaxies at
$z\approx 0$, as determined by Sheth et al. (2003),
\begin{eqnarray}
\nonumber \frac{dN}{d\sigma_v}d\sigma_v&=&1.34\times 10^{-3}
(h_{70}^{-1}\,{\rm Mpc})^{-3} \left({\sigma_v\over88.8{\rm
km\,s}^{-1}}\right)^{6.5}\\
&&\times{\rm\,exp}\left[-\left({\sigma_v\over 88.8{\rm
km\,s}^{-1}}\right)^{1.93}\right]\frac{d\sigma_v}{\sigma_v}.
\label{eq:dndsigma}
\end{eqnarray}
This function, shown by the dashed curve in Figure~\ref{fig:dndsig},
shows a clear peak: the abundance of early type galaxies declines at
velocity dispersions below $\lsim 200~{\rm km~s^{-1}}$.  Below this
value, the population of spheroids is dominated by the bulges of
late--type galaxies, as demonstrated by the solid curve showing the
sum of the two spheroid velocity functions.  For reference, the
data--points in the figure show the observed quasar LF at redshift
$z=2.18$ (adapted from Hopkins et al. 2006), using the mean $L_{\rm
Q}-\sigma_v$ relation from equation~(\ref{eq:Lsigma}) without any
scatter. It is clear from the figure that the quasar LF at faint
luminosities is monotonic, and traces the total spheroid population
much more closely than the subset represented by the early type
galaxies.  The empirical evidence listed in the Introduction also
suggests that the correlation between SMBH mass and spheroid
properties apply both to SMBHs in early type galaxies and to those in
bulges of late--type galaxies; it is therefore important to include
late--type bulges in our analysis.  Unfortunately, we do not have a
direct measurement of the age--distribution of these bulges as a
function of their velocity dispersion.  However, bulges lie very close
to the fundamental plane of elliptical galaxies, and appear to have
similar formation epochs (Falc\'on-Barroso et al. 2002); we therefore
make the reasonable assumption that the age--distribution of a bulge
is identical to that of an early--type galaxy with the same velocity
dispersion.  Furthermore, we follow Sheth et al. (2003) and identify
$\sigma_v=V_c/\sqrt{2}$ as the bulge velocity dispersion for a
late--type galaxy with circular velocity $V_c$. This identification is
justified by a recent, more detailed population synthesis study (Tundo
et al. 2006).

We next assume that the stars in each of the nearby spheroid formed in
a single burst, with the age $t_e$ identified as the mass--weighted
stellar age, inferred from the spectral fits of early--type galaxies
discussed above.  Both the mean age, and the shape of the
age--distribution, are found to be a strong function of the velocity
dispersion $dp/dt_e=dp/dt_e(\sigma_v, t_e)$ (see
Figure~\ref{fig:agevel} above).

We also assume that the velocity dispersion $\sigma_v$ remains
constant after the formation of the galaxy.  Dry mergers between
gas--poor early type galaxies may indeed cause only a modest change to
$\sigma_v$, as expected based on idealized models and confirmed by
numerical simulations (e.g. Nipoti et al 2003, although the precise
conclusions depend on the merging galaxies' structure and their
orbital parameters; Robertson et al. 2006, Boylan-Kolchin et
al. 2006).  If the BH masses add linearly, then such dry mergers
cannot maintain the $M_{\rm bh}-\sigma_v$ relation, implying that
mergers are either infrequent, or that they are accompanied by
dissipation.  Furthermore, in the high--resolution N--body simulations
of Zhao et al. (2003a,b), dark matter halos grow in two phases. An
initial rapid merger of many progenitors with comparable masses build
up the potential well; this is followed by a slower accretion of
lower--mass satellites adding mass primarily to the outskirts of the
halo, without affecting the central structure.  Our modeling
implicitly requires only that most of the BH mass is assembled
(possibly in multiple mergers) within a small fraction of the Hubble
time.

Under the assumption that a SMBH forms concurrently with the stars in
each early--type galaxy, we then adopt the SMBH mass given by the
relation,
\begin{equation}
\log{\left(\frac{M_{\rm bh}}{\rm M_\odot}\right)}=
\alpha+\beta\log{\left(\frac{\sigma_v}{200\mbox{km\,s}^{-1}}\right)}
\label{eq:Msigma}
\end{equation}
with $\alpha=8.15$ and $\beta=3.83$. This is the mean relation
obtained by Tundo et al. (2006) while allowing an intrinsic scatter in
$M_{\rm bh}$ at fixed $\sigma_v$ (determined to be $\sim 0.2$ dex).
Although current SMBH samples suffer from selection bias, Bernardi et
al. (2006b) suggest that the $M_{\rm bh}-\sigma_v$ relation is
unlikely to have been strongly affected.  Wyithe (2006) recently found
that an additional quadratic term in the right--hand side of
equation~(\ref{eq:Msigma}), $\beta_2[\log(\sigma/200{\rm
km~s^{-1}})]^2$, is required.  Although the statistical significance
of a nonzero $\beta_2>0$ is modest ($\approx 1-1.5\sigma$ depending on
the sample used), we have repeated our analysis below using the
best--fit model in Wyithe (2006), with $(\alpha,\beta,\beta_2)=(8.05,
4.2, 1.6)$, and have verified that our predictions for the quasar LF
change by much less than their uncertainties.

We assume that the SMBHs gain most of their mass in a luminous
short--lived phase, at the end of their rapid assembly.  During this
phase, they radiate at some fraction $f_{\rm Edd}$ of the Eddington
limit of their final mass, for a duration $t_{\rm Q}$ years. During
this time, the SMBH is identified as a luminous quasar with a fixed
luminosity $L_{\rm Q}$.  We expect the lifetime to be of order the
Salpeter time, $t_{\rm Q}\sim 4\times 10^7$yr (Martini 2004; see also
Shankar et al. 2004).  Note that this lifetime is identified as the
cumulative duration of quasar activity (i.e., if the activity is
episodic, then it represents the sum of all episodes), and also
assumes that the quasar is unobscured by gas or dust during the epoch
of bright activity.  This latter assumption may be justified in the
scenario in which the lifetime is determined by feedback: the bright
quasar epoch corresponds to an expulsion of the gas and dust from
around the nucleus (which removes the obscuration on one hand, and
terminates quasar activity on the other).  This type of model of
quasar activity has been proposed recently by Hopkins et al. (2005,
and references therein), although we caution that in their model, the
BHs spend a longer time in an obscured phase (see discussion in
\S~\ref{sec:results} below).

While it is useful to think in terms of separate $M_{\rm bh}-\sigma_v$
and $L_{\rm Q} - M_{\rm bh}$ relations (the latter specified by
$f_{\rm Edd}$), both of which have been empirically constrained in
various galaxy samples, our model predictions depend only on the
combination of these two, i.e on the dependence of the quasar's
luminosity $L_{\rm Q}$ on $\sigma_v$.  We therefore introduce the
parameter $\alpha_L$ giving the normalization of the $L_{\rm
Q}-\sigma_v$ relation for a SMBH in a host with velocity dispersion
200 ${\rm km s^{-1}}$.  We then have the bolometric luminosity
\begin{equation}
\log{\left(\frac{L_{\rm Q}}{\rm L_\odot}\right)}=
\alpha_L+12.72+\beta\log{\left(\frac{\sigma_v}{200\mbox{km\,s}^{-1}}\right)},
\label{eq:Lsigma}
\end{equation}
where $10^{4.57} {\rm L_\odot/M_\odot}$ is the Eddington luminosity,
so that $\alpha_L=0$ corresponds to the Eddington luminosity of a
$10^{8.15} {\rm M_\odot}$ black hole, i.e. to the adopted
normalization of the $M_{\rm bh}-\sigma_v$ relation and $f_{\rm
  Edd}=1$.  In our analysis, we allow the normalization $\alpha_L$ to
vary with redshift; however, we keep the logarithmic slope fixed,
$d\log L_{\rm Q}/d\log\sigma_v=\beta=3.83$.

Under the above assumptions, and in the limit of a negligible scatter
in $L_{\rm Q}$ at fixed $\sigma_v$, the space density of luminous
quasars at redshift $z$ is predicted to be
\begin{eqnarray}
\frac{dN^{0}}{dL_{\rm Q}}(z)=
\left(\frac{dN}{d\sigma_v} \right)
\left[t_{\rm Q}\times \frac{dp}{dt_e}(\sigma_v,t_e)\right]
\left(\frac{dL_{\rm Q}}{d\sigma_v}\right)^{-1}.
\label{eq:dndl}
\end{eqnarray}
In the above equation, $t_e$ is the lookback time to redshift $z$,
$dp/dt_e$ is unresolved and assumed to be a constant in each age bin
of width $dt_e=1$Gyr, and $L_{\rm Q}$ is the bolometric luminosity
given by equation~(\ref{eq:Lsigma}).

Since the $M_{\rm bh}-\sigma_v$ relation and $f_{\rm Edd}$ are both
likely to have intrinsic scatter, we allow a probability distribution
$dp/dL_{\rm Q}$ of quasar luminosities at fixed $\sigma_v$. To be
specific, we assume that $dp/dL_{\rm Q}$ follows a log--normal
distribution, i.e. $dp/d\log L_{\rm Q}$ is described by a
Gaussian,\footnote{Note that here and elsewhere in the paper, $\log$
  refers to base 10 logarithm.} with a width independent of
$L_{\rm Q}$.  This can be achieved by allowing $\alpha_L$ to have a
Gaussian distribution with a mean $\langle \alpha_L \rangle$ and
standard deviation $\Delta \alpha_L$.\footnote{Note that with this
  definition $\langle L_{\rm Q}\rangle \neq
  10^{\langle\alpha_L\rangle}$.  Our definition is consistent with
  Tundo et al. (2006), i.e. equation~(\ref{eq:Msigma}) describes the
  logarithmic mean $\langle \log M_{\rm bh} \rangle$.}
$\Delta \alpha_L$ can therefore
be regarded as the combined scatter in the $M_{\rm bh}-\sigma_v$
relation (at fixed $\sigma_v$) and in $f_{\rm Edd}$ (at fixed $M_{\rm
  bh}$).  To predict the quasar LF in the presence of scatter, the
quantity in equation~(\ref{eq:dndl}) has to be convolved with
$dp/dL_{\rm Q}$,
\begin{eqnarray}
\nonumber
\frac{dN}{d\log L_{\rm Q}}&=&
\int_{0}^{\infty} d\log L_{\rm Q}^0
\frac{dN^{0}}{d\log L_{\rm Q}^0}
\left. \frac{dp}{d\log L_{\rm Q}}\right|_{\log L_{\rm Q}^0}\\
&=&
\int_{0}^{\infty} d\log L_{\rm Q}^0
\frac{dN^{0}}{d\log L_{\rm Q}^0}
\frac{dp}{d\alpha_L}.
\label{eq:dndlscat}
\end{eqnarray}
In these equations, $\log L_{\rm Q}^0$ represents a mean logarithmic
luminosity, computed from equation~(\ref{eq:Lsigma}) with
$\alpha_L=\langle\alpha_L\rangle$, and in the last equation, the
Gaussian $dp/d\alpha_L$ is evaluated at $\alpha_L=\log L_{\rm Q}-\log
L_{\rm Q}^0$.

In summary, the parameters of the model described above are $t_{\rm
  Q}$, $\langle \alpha_L \rangle$, and $\Delta \alpha_L$, describing
the duration of quasar activity, and the mean luminosity and its
scatter during the quasar epoch of a SMBH in a spheroid with fixed
$\sigma_v$.  We will fit these three parameters simultaneously and
independently in each redshift bin, i.e. we do not require ab--initio
that these parameters be redshift--independent.  We note again that
for comparison to previous work, it can be useful to regard the
$L_{\rm Q}-\sigma_v$ relation we use in our fits as the combination of
an $M_{\rm bh}-\sigma_v$ relation and the Eddington ratio $f_{\rm
  Edd}$.

\section{Comparing Predicted and Observed Quasar LFs}
\label{sec:stat}

\begin{figure}[t]
\plotone{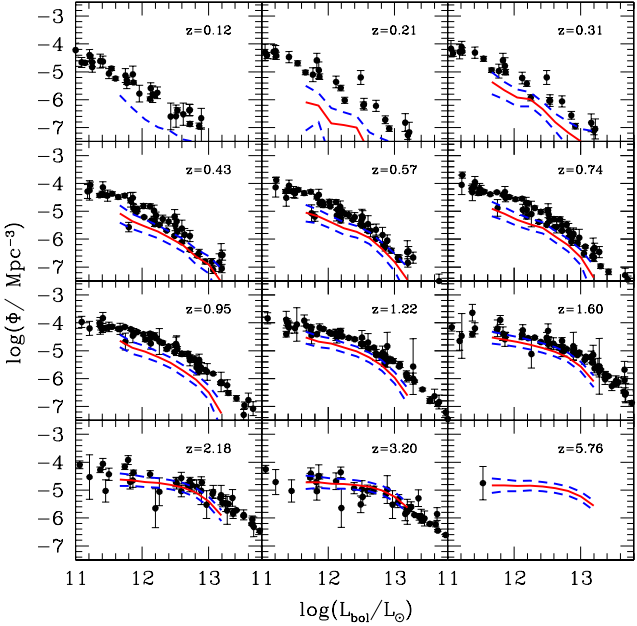}
\caption{This figure shows the predicted and observed quasar LF at 12
  different redshifts, as labeled (corresponding to look--back times
  of 1.5--12.5 Gyr, in increments of 1 Gyr).  The solid (red) curves
  follow from a model with $L_{\rm Q}-\sigma_v$ normalization
  $\alpha_L=0$ (which corresponds to $f_{\rm Edd}\equiv L_{\rm
    Q}/L_{\rm Edd}=1$ with our adopted $M_{\rm bh}-\sigma_v$
  normalization).  The model assumes no scatter, and a quasar phase
  that lasts for a Salpeter time at a radiative efficiency of 10
  percent ($t_{\rm Q}=4\times10^7$ years).  The dashed (blue) curves
  indicate the uncertainty in the model prediction, arising from
  Poisson sampling, age--determination, and local galaxy abundance
  errors (in the top left panel, only the upper limit is visible).
  The black data--points have been adapted from a recent compilation
  by Hopkins et al. (2006).  }
\label{fig:LF0}
\end{figure}

Our main task is to compare the predictions of the above model to the
observed quasar LF.  The bolometric quasar LF has been recently
inferred from a compilation of multi--wavelength data (Hopkins et
al. 2006).  Here we adopt the observational data--points, with their
quoted statistical errors, as reported in this compilation, but
re--binned to correspond to the LF in 12 different redshift intervals,
centered at look--back times of $t=1.5-12.5$ Gyr, in increments of 1
Gyr.  Each redshift bin has a width corresponding to $\Delta t = \pm
0.5$ Gyr.  These data, together with their errors, are shown in
both Figures~\ref{fig:LF0} and \ref{fig:LF}.

In order to compare our model predictions to these data, several
additional uncertainties need to be taken into account.  In
determining the fraction of galaxies, at fixed $\sigma_v$, that fall
within a given age bin, there are two main sources of errors.  First,
there is a Poisson sampling error $\delta n_{\rm p}$, which is
significant in those age-- and $\sigma$--bins that have only a few
galaxies (the number of galaxies is indicated by the label in each
panel in Fig.~\ref{fig:agevel}).  Second, errors in the age
determinations (increasing from $\delta t_e < 1$ Gyr for $t_e\lsim 7$
Gyr to $\delta t_e \approx 3$ Gyr for $t_e\gsim 10$ Gyr) can shift
galaxies between age--bins.  We estimated this error, $\delta n_{\rm
  t}$, by performing 20 Monte Carlo realizations of each of the
age--distributions, generating a random age of each galaxy within its
allowed range returned by MOPED.  We maintain the prior that the age
of any galaxy, when extrapolated to $z=0$, cannot exceed 13.7 Gyr, and
identify $\delta n_{\rm t}$ as the range that includes 14 out of the
20 realizations.  The resulting uncertainties $\delta n_{\rm t}$ are
shown by the error--bars in Figure~\ref{fig:agevel}.  We add $\delta
n_{\rm p}$ and $\delta n_{\rm t}$ in quadrature, to obtain the total
uncertainty in the fraction of galaxies falling in each age--bin.

In obtaining the total (normalized) space density of quasars, there is
an additional uncertainty, from the measurement errors of the local
velocity function of early type galaxies (Sheth et al. 2003).  The
statistical errors are small ($\lsim 15\%$, although they increase
toward the bright end), and systematic errors, due to
including/excluding low S/N galaxies, are somewhat larger, 20-25\%.
We chose to include an additional, independent 15\% error in each
luminosity bin in the quasar space density, although we note that in
practice, these errors are much smaller than the combination of the
other two sources of error and the observational errors, discussed in
the preceding paragraphs.

Finally, there is additional uncertainty in the prediction of the
quasar LF at the faint end (corresponding to the threshold $\sigma_v
\approx 100$ km/s) in those models with a large scatter ($\Delta
\alpha_L\gsim 0.5$).  In these models, the velocity function, and the
age distributions, have to be extrapolated to $\sigma_v < 100$ km/s.
While the uncertainty introduced by this extrapolation cannot be
rigorously quantified, we include it in our error budget as follows.
We first compute the quasar LF in the absence of scatter, $N^0_{\rm
  Q}\equiv L_{\rm Q}(dN^0/dL_{\rm Q})$, by extrapolating the
power--law slope implied by equation~(\ref{eq:dndl}), and convolve
$N^0_{\rm Q}$ with the given scatter $\Delta \alpha_L$ following
equation~(\ref{eq:dndlscat}). We then repeat this calculation, but
reducing/increasing $dp/dt_e$ in each $t_e$ bin by its $1\sigma$
uncertainty as discussed above.  This results in a flatter/steeper
slope for $N^0_{\rm Q}$, and yields a lower/higher $N_{\rm Q}$. We
adopt these values as the $1\sigma$ lower/upper limits on $N_{\rm Q}$.
In practice, this increases the Poisson error on $\Delta N_{\rm Q}$
significantly in models with scatter relative to those with no
scatter.  As an example, at $z=0.31$, at the faintest $L$ the
uncertainty is a factor of $\sim 3$ without scatter (shown in
Fig.~\ref{fig:LF0}), but is about a factor of $\sim 6$ for the model
with $\Delta \alpha_L=0.6$ (shown in Fig.~\ref{fig:LF}).

In summary, given a set of parameters, $t_{\rm Q}$, $\langle \alpha_L
\rangle$, and $\Delta \alpha_L$, we compute the expected space density
of quasars in each logarithmic luminosity bin, $N_{\rm Q}$ as
described in the previous section, and at each redshift and
luminosity, we compute the error $\Delta N_{\rm Q}$ by adding in
quadrature the Poisson sampling errors, the errors caused by the
age--determinations, and the uncertainties in the local galaxy
abundance.  We then compute a $\chi^2$ for each data--point in the
Hopkins et al. (2006) compilation that falls within the predicted
range of luminosities.  Here $\chi^2$ is defined as the ratio $\Delta
N_{\rm Q, diff}/\Delta N_{\rm Q, err}$, where $dN_{\rm Q, diff}$ is
the difference between the predicted and measured quasar space
densities, and $dN_{\rm Q, diff}$ is the sum in quadrature of the
modeling errors and the observational errors discussed above.  Note
that the number of data--points available for the comparison of
predictions vs. observations (and hence the number of degrees of
freedom in the fit) varies with $\langle \alpha_L \rangle$.  This is
because $\sigma_v$ lies in the range $100~{\rm km~s^{-1}} \lsim
\sigma_v \lsim 280~{\rm km~s^{-1}}$, but the corresponding range of
$L_{\rm Q}$ shifts with $\alpha_L$.

At each redshift, the above procedure yields a total $\chi^2$, summed
over the available observational data--points, as a function of
$t_{\rm Q}$, $\langle \alpha_L \rangle$, and $\Delta \alpha_L$.  As a
single figure of merit for the goodness--of--fit of a model, we
further sum the total $\chi^2$ over all of the 12 redshift bins, and
obtain a cumulative reduced $\chi^2$ (where the number of degrees of
freedom, i.e. the number of data--points, minus 3 for the 3
independent fitting parameters, is typically of order $\sim 100$).

\section{Results and Discussion}
\label{sec:results}

\begin{figure}[t]
\plotone{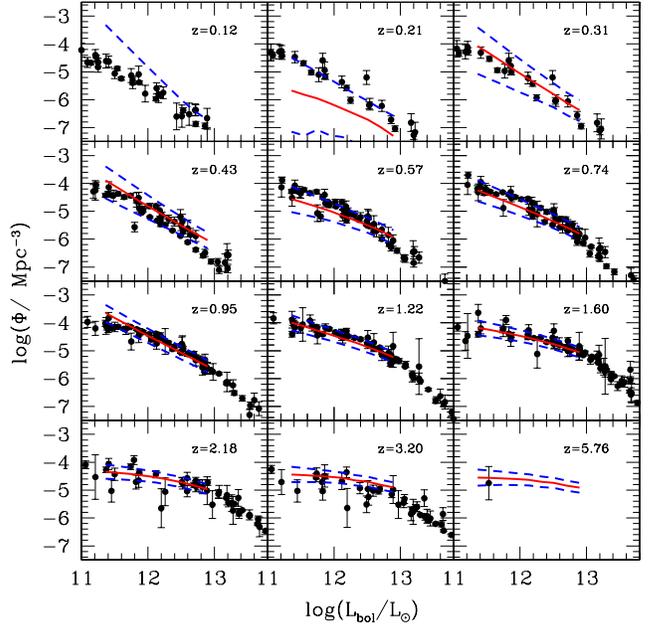}
\caption{This figure shows the predicted and observed quasar LF at 12
  different redshifts, together with their uncertainties, as in
  Figure~\ref{fig:LF0}.  The curves here assume $\alpha_L=-0.3$
  (corresponding to a constant, redshift--independent mean Eddington
  ratio of $\langle \log f_{\rm Edd}\rangle=-0.3$ with our adopted
  $M_{\rm bh}-\sigma_v$ normalization), a scatter of 0.6 dex in
  $L_{\rm Q}$ at fixed $\sigma_v$, and a quasar lifetime of $8\times
  10^7$ years.  This model a statistical fit with a reduced $\chi^2$
  of 1.1 (see text for discussion), for the entire redshift range as a
  whole. Note that the presence of scatter increases the uncertainties
  relative to Figure~\ref{fig:LF0}.}
\label{fig:LF}
\end{figure}

In Figures~\ref{fig:LF0} and \ref{fig:LF}, we show the predicted
quasar LF at 12 different redshifts, compared to the observed
bolometric LF adapted from the compilation by Hopkins et al. (2006;
shown as black points with error--bars).  The twelve different panels
correspond to 1 Gyr--wide bins in look--back times, centered on
$t_e=1.5-12.5$ Gyr, in increments of 1 Gyr, and to redshifts as
labeled.  The curves in Figure~\ref{fig:LF0} have been obtained in a
``basic'' model with the most na\"{\i}ve set of assumptions, namely
$\alpha_L=0$ with no scatter, and a quasar phase lasting for a
Salpeter time at a radiative efficiency of 10 percent ($t_{\rm
  Q}=4\times10^7$ years).  This model is shown for illustration. The
solid (red) curves show the predicted quasar LF, and the dashed (blue)
curves bracket the estimated $1\sigma$ uncertainties, as discussed in
\S~\ref{sec:stat} above.

As the figure shows, this na\"{\i}ve model under--predicts the overall
normalization of the LF (especially at low redshifts), and predicts a
shape that declines too steeply at the bright end (especially at high
redshift). Therefore, under the most naive assumptions, the
age-distributions we infer for the early-type populations do not
``map'' the local population of early-type galaxies onto the evolving
quasar population.  The age-distributions have the right property, in
that larger early-types are older (as inferred by numerous previous
works on ``down-sizing'').  This is in accord with the increase in the
characteristic quasar luminosity toward high $z$ (e.g. Cattaneo \&
Bernardi 2003; Haiman, Ciotti \& Ostriker 2004). However, over the
range of velocity dispersions $100~{\rm km~s^{-1}} < \sigma_v <
280~{\rm km~s^{-1}}$ for which we can determine the age--distribution,
the corresponding quasar LF flattens considerably from $z=0$ to
$z=2-3$.  This flattening is not reproduced by the dependence of the
observed age--distribution on $\sigma_v$.

We have searched through the full 3--dimensional parameter space
($t_{\rm Q}$, $\langle \alpha_L \rangle$, $\Delta \alpha_L$) for a set
of values that corrects these deficiencies.  Changes in $t_{\rm Q}$
and $\langle\alpha_L\rangle$ amount to shifting the predicted
LF along the vertical or horizontal axes in Figure~\ref{fig:LF},
respectively, without changing the predicted shape.  An increased
scatter (higher $\Delta \alpha_L$) also shifts the LF up along
the vertical axis, but it also flattens the predicted LF, by
preferentially increasing the abundance at the bright, steep tail.

The solid (red) curves in Figure~\ref{fig:LF} show predictions from a
model with parameter choices of ($t_{\rm Q}$, $\langle
\alpha_L\rangle$, $\Delta \alpha_L$) = ($8\times 10^7$ years, -0.3,
0.5).  This model has a reduced $\chi^2$ of 1.1 ($\chi^2=106$ for 98
degrees of freedom over the twelve redshift bins).  The dashed curves
indicate the uncertainty in the model predictions in each age bin.
The uncertainty increases at the faint end of the LF, and at lower
redshifts (as well as in the highest $z$ bin), as there are only a
handful of SDSS galaxies in these velocity dispersion and age bins
(see Figure~\ref{fig:agevel}).  As noted above, the predicted LF at
the faint end has an additional uncertainty when the scatter is large,
since the prediction involves extrapolating to $\sigma_v < 100$ km/s.

\begin{figure}[t]
\plotone{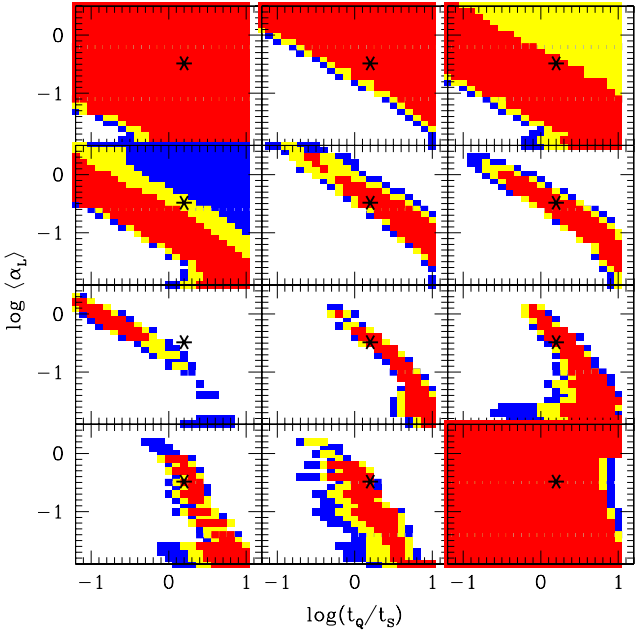}
\caption{This figure shows, in each redshift bin, contours of constant
  reduced $\chi^2$ in that bin. The blue/yellow/red contours
  correspond to $\chi^2=2.0, 1.5$, and $1.0$.  The x-axis in each
  panel shows the quasar lifetime in units of the Salpeter time,
  $4\times10^7$ yr, and the y-axis shows the mean $\alpha_L$
  (equivalent to the mean Eddington ratio with our adopted $M_{\rm
  bh}-\sigma_v$ normalization). The scatter in all models are fixed at
  $\Delta \alpha_L=0.9$.}
\label{fig:probs}
\end{figure}

This model, while simple, is in reasonable overall agreement with the
observational data.  There does not appear to be compelling evidence
that either the duty cycle or the Eddington ratio (or $M_{\rm
  bh}-\sigma_v$ relation) evolves significantly with redshift.  
%As the
%figure shows the model slightly overpredicts the number of
%quasars at $z\approx 1$ at the faintest luminosities.  This
%overprediction can be traced directly to the sharp feature between
%$7-8$ Gyr in the age distributions for the smallest galaxies shown in
%Figure~\ref{fig:agevel}: the sharp rise translates to an increase in
%the predicted quasar abundance from $z=0.74$ to $z=0.95$.  This
%feature, however, is unlikely to be physical.  
It is also worth pointing out that the quasar lifetime preferred by
this model is about a factor of two longer than the fiducial Salpeter
time of $4\times10^7$ years (at a radiative efficiency of
$\epsilon=0.10$).

In Figure~\ref{fig:probs}, we assume a larger scatter of $\Delta
\alpha_L=0.9$, and we show contours of constant reduced $\chi^2$ as a
function of the other two parameters. The reduced $\chi^2 $ was
computed separately in each of the redshift bins (each bin having
$\sim$ 10 d.o.f. on average).  The blue/yellow/red contours correspond
to $\chi^2=2.0, 1.5$, and $1.0$, respectively, and the star in each
panel marks the model that best fits when all 12 redshift bins are
combined, ($t_{\rm Q}$, $\langle \alpha_L\rangle$, $\Delta \alpha_L$)
= ($6.3\times 10^7$ years, -0.5, 0.9).  As this figure shows, no
combination of parameters can match the observed LF at the precision
of $\chi^2<1.5$ simultaneously at all redshifts (the worst fit is at
$z=0.95$, as we discussed above).  On the other hand, the figure shows
significant degeneracies: roughly, a longer lifetime can be ``traded''
for a smaller $\alpha_L$ (or equivalently, a smaller Eddington ratio).
This is expected: a smaller $\alpha_L$ shifts our predicted LF curves
to the left (a given $L_{\rm Q}$ then corresponds to a larger
$\sigma_v$, i.e. a smaller $dn/d\sigma_v \times dp/dt_e$), while a
longer lifetime moves the curves upward (scaling linearly).  This
explains the sense of the degeneracy for a given $L_{\rm Q}$.
Likewise, we found (not shown on the figure) that an increased scatter
can be compensated by a shorter lifetime and/or a smaller Eddington
ratio.

Overall, there is a suite of models that can fit the LF over most of
the redshift range.  In particular, models along lines with $\log
t_{\rm Q}+\langle \alpha_L\rangle \approx$ constant are degenerate,
with models with longer lifetimes only mildly disfavored.  For
example, the models with ($t_{\rm Q}$, $\langle\alpha_L\rangle$,
$\Delta \alpha_L$) = ($6.3\times 10^7$ years, -0.5, 0.9) or
($1.6\times 10^8$yr, -1.6, 1.2) both have overall reduced $\chi^2\lsim
1.0$ (with all twelve redshift bins combined).  We find that models
with even larger scatter can have even smaller reduced $\chi^2$; this
is a result of the large prediction uncertainty in these models. Such
large--scatter models, with somewhat longer lifetimes, however,
require exceedingly small $\alpha_L$, or equivalently, small Eddington
ratios ($\lsim 0.03$), and are disfavored by other considerations (see
discussion below).

\begin{figure}[t]
\plotone{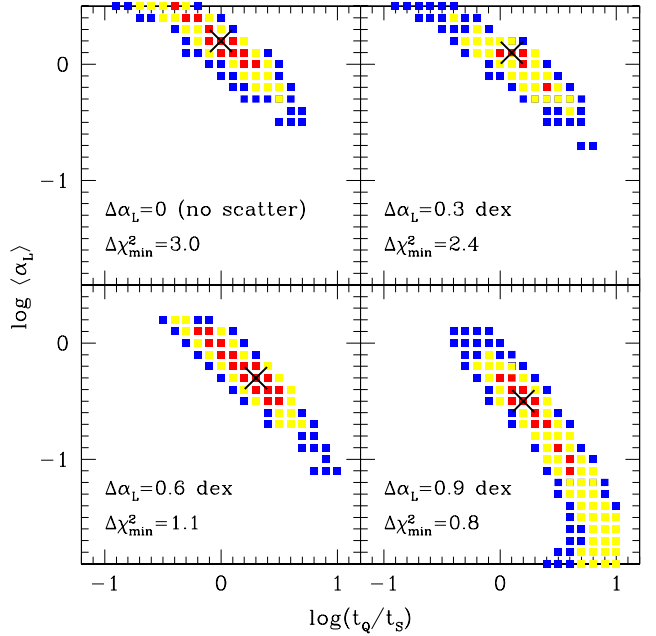}
\caption{This figure shows contours of constant reduced $\chi^2$
  combined over all 12 redshift bins. The scatter is fixed at four
  different values ($\Delta\alpha_L=0, 0.3, 0.6, 0.9$). In each panel,
  the large cross marks the best--fit model. Note that the best--fit
  models have smaller reduced $\chi^2$ when the scatter is
  increased. The contour levels in each panel are adjusted to the
  minimum of $\chi^2$, and are different in each panel. The
  blue/yellow/red contours enclose points with $\chi^2$ below (4.0,
  5.5, 8.0), (3.0, 5.0, 7.0), (2.0 3.0 4.0), and (1.0 1.5 2.0) in the
  four panels, respectively (from top left, clockwise).  These
  contours serve to illustrate the direction of the degeneracies
  discussed in the text, and show that a non--negligible scatter ($\Delta
  \gsim 0.6$) is required.}
\label{fig:totalprobs}
\end{figure}

In Figure~\ref{fig:totalprobs}, we show contours of constant reduced
$\chi^2$ with all 12 redshift bins combined.  The four panels fix the
scatter at four different values ($\Delta\alpha_L=0, 0.3, 0.6, 0.9$).
In each panel, the large cross marks the best--fit model. As explained
above, the models with large scatters have best--fits with lower
reduced $\chi^2$, shown in each panel.  As the figure shows,
$\Delta\alpha_L < 0.6$ yields poor fits. We therefore find that the
fits require a non--negligible scatter; this is a result of the quasar
LF being flatter at the bright end then the velocity--function of
spheroids (as seen clearly in Figure~\ref{fig:dndsig}).  The contour
levels in each panel are adjusted to the minimum of $\chi^2$, and are
different in each panel. The blue/yellow/red contours enclose points
with $\chi^2$ below (4.0, 5.5, 8.0), (3.0, 5.0, 7.0), (2.0 3.0 4.0),
and (1.0 1.5 2.0) in the four panels, respectively (from top left,
clockwise).  These contours serve to illustrate the direction of
the degeneracies discussed above.

We note that our conclusions above are based on the mass--weighted
stellar age as a proxy for the time the SMBH forms.  The
mass--weighted age is likely to be the appropriate estimate, if the
bulk of the stars and the SMBH forms concurrently.  However, to check
the sensitivity of our results to this assumption, we have repeated
our analysis, replacing the mass--weighted age either by a
luminosity--weighted age, or an alternative age--estimate based on the
(H$\gamma_{\rm F}$) absorption line index.  In both cases, we find
that the ages are decreased (the decrease is more pronounced when the
absorption-line ages are used). This increases the predictions for the
quasar LF at low redshifts ($z\lsim 1$), and, as a result, we find
that matching the observed LF would require that either
$\langle\alpha_L\rangle$, $\Delta \alpha_L$, and/or $t_{\rm Q}$
decline from $z\approx 2$ to $z=0$ by a factor of a few.  One or more
of the three model parameters would therefore have to evolve to fit
the quasar LF over the entire range $0<z\lsim 6$.

Our result also assumes that the luminous quasar phase is short, and
it follows promptly the formation of the SMBH.  The accretion history
and the corresponding ``lightcurve'' of SMBHs can be more extended and
complex. We first note that in the best--fit models we find with
reduced $\chi^2\sim 1$, the quasar lifetime is a factor of $1.5-2$
longer than the Salpeter time, while the mean Eddington ratio is
$0.3-0.5$.  At fixed radiative efficiency, this implies that during
the quasar phase, the BH mass, and the accompanying luminosity,
increases by a factor of $\sim 2$. In our models, we ignored this
increase, and assumed a constant $L_Q$ during the luminous phase.  In
practice, such a variation in $L_{\rm Q}$ would mimic a smaller
Eddington ratio and a larger scatter in our models. Given the
uncertainties about the actual quasar light--curve, we chose not to
model this variation here.

Observations of X--ray spectra (e.g. Wall et al. 2005; Barger et
al. 2005) suggest that a SMBHs can suffer significant obscuration, an
effect that increases at lower luminosities.  For example, Cattaneo \&
Bernardi (2003) included a luminosity-dependent obscuration in their
predictions for the quasar LF.  In their model, the evolution of
bright quasars is determined by the age-distribution, but the
evolution of faint quasars is driven by the assumed form of the
obscuration (which reduces the abundance of faint quasars).  They
showed that the obscuration can achieve the required flattening of the
LF (which, in our case, is attributed to scatter instead).  On the
other hand, the X--ray observations, and also theoretical arguments
(Hopkins et al. 2005, 2006) suggest that SMBHs spend a significantly
longer period in their obscured phase, during which they shine well
below the Eddington luminosity.  Such a luminosity--dependent lifetime
would further change the shape of the predicted LF, and would likely
require the introduction of a more detailed model, in which $t_{\rm
Q}$, $\langle \alpha_L \rangle$ and $\Delta \alpha_L$ depend on the
SMBH mass.

Taken at face value, however, our results imply that the duty cycle,
the Eddington ratio, and the $M_{\rm bh}-\sigma_v$ relation, do not
necessarily evolve with redshift.  Our results do not, conversely,
imply that evolution is ruled out. For example, one could adopt
different points from the degenerate innermost red contours shown in
Figure~\ref{fig:LF0} at each redshift. Alternatively,
Figure~\ref{fig:totalprobs} shows that the best-fit values of
$\langle\alpha_L\rangle$ and $t_Q$ do change as scatter is varied; one
could therefore postulate an evolving $\langle\alpha_L\rangle$ (say,
decreasing with redshift) together with an evolving scatter (which
would have to then increase with redshift).

There have been suggestions in the recent literature that the
normalization of the $M_{\rm bh}-\sigma_v$ relation relation may
increase toward high $z$, with $M_{\rm bh}$ larger at fixed $\sigma_v$
(e.g. Shields et al. 2006; Wandel 2004) -- by as much as a factor of 4
at $z=0.36$ than at $z=0$ (Woo et al. 2006), although these results
could arise from a selection effect of preferentially missing the
lower luminosity BHs at high $z$, or the high luminosity BHs at low
$z$ (Bernardi et al. 2006b).

Eddington ratios are difficult to determine empirically (although this
could be possible in the future with high precision, if gravity waves
can be detected by LISA and identified with luminous quasars; Kocsis
et al. 2006).  Kollmeier et al. (2006) combined measurements of line
widths and continua of a sample of AGN at $0.3<z<4$ to infer $\langle
f_{\rm Edd}\rangle \sim 0.3$, with a scatter of $\sim 0.3$dex, with no
apparent evolution in the mean or the scatter (note that if the
fueling rate falls below the Eddington rate, it would be surprising if
it had no scatter).  Vestergaard (2004) estimated Eddington ratios in
a sample of high redshift quasars using an observed correlation
between the size of the broad line region and the luminosity of the
quasar (the correlation is calibrated using reverberation mapping of
lower redshift objects; e.g. Kaspi et al. 2000; Vestergaard
2002). They find values ranging from $\approx 0.1$ to $\gsim 1$, with
the $z~\gsim 3.5$ quasars having a somewhat higher mean $L/L_{\rm
Edd}$, and a narrower distribution, than the lower redshift
population.  In a more extensive lower redshift $0<z<1$ sample, Woo \&
Urry (2002) also find higher Eddington ratios towards $z=1$. However,
these results may represent a trend towards higher ratios at higher
luminosities, and could also be explained by selection effects.
Whether there exists any trend, and whether it is primarily with
redshift or luminosity is an important question, but large scatter and
selection effects presently preclude a firm answer.

Our fits are in good agreement with these Eddington ratio
measurements.  Our results suggests that, in order to flatten the
predicted quasar LF at $z=2-3$, a scatter {\it must} be present in the
relation between $L$ and $\sigma_v$.  Tundo et al. (2006) find an
intrinsic scatter of $\approx$ 0.22 dex, while Wyithe (2006) and
Marconi and Hunt (2004) both find 0.3 dex, in the $M_{\rm
bh}-\sigma_v$ relation, and Lapi et al. (2006) find a similar scatter
of 0.3 dex in $M_{\rm BH}$ at fixed virial mass.  The combined scatter
in the $M_{\rm bh}-\sigma_v$ relation (0.2-0.3 dex) and in $f_{\rm
Edd}$ (0.3 dex from Kollmeier et al.), when added in quadrature, is
$\sim 0.4$ dex, which is in fair agreement with our finding that
$\gsim 0.6$ dex is required (for a reduced $\chi^2$ of 1.1 or smaller,
see Figure~\ref{fig:totalprobs}).  However, it is interesting to note
that our minimum scatter is slightly larger than the combined scatter
in the $M_{\rm bh}-\sigma_v$ relation and $f_{\rm Eddd}$, and a factor
of 3 larger than the scatter in the $L_*-sigma_v$ relation of
early-type galaxies, which is $\sim 0.2$dex.

\section{Conclusions}
\label{sec:conclude}

In this paper, we attempted to predict the evolution of the optical
quasar luminosity function, over the redshift range $0<z\lsim 6$, in a
simple model with three free parameters, which is based on the
inferred age--distribution of early type galaxies, and makes the
assumption that quasar black hole growth and star--formation track one
another at all times.  Our ``best--fit'' models have quasar lifetimes
($\approx 6-8\times 10^7$ yr), mean Eddington ratios ($\approx
0.3-0.5$) that are in good agreement with other determinations. Our
results also require a non--negligible scatter between the velocity
dispersion of the galaxy and the mass of its resident BH ($\approx
0.6-0.9$ dex).  Our basic conclusion is that the quasar LF can be fit
to satisfactory accuracy in these simple models, without a compelling
need for any of the model parameters to evolve with redshift between
$0<z\lsim 6$.  This result supports the direct connection between the
build--up of spheroids and their nuclear SMBHs.

\vspace{-0.5\baselineskip}

\acknowledgements{We are grateful to Philip Hopkins for providing the
  re--binned data--points for the quasar LF in Figure~\ref{fig:LF},
  and an anonymous referee whose comments helped to significantly
  improve this paper. ZH acknowledges partial support by NASA through
  grants NNG04GI88G and NNG05GF14G, by the NSF through grants
  AST-0307291 and AST-0307200, and by the Hungarian Ministry of
  Education through a Gy\"orgy B\'ek\'esy Fellowship. RJ is partially
  supported by NSF grant PIRE-0507768 and NASA grant NNG05GG01G. MB is
  partially supported by NASA grant LTSA-NNG06GC19G, and by grants
  10199 and 10488 from the Space Telescope Science Institute, which is
  operated by AURA, Inc., under NASA contract NAS 5-26555. }

%\vspace{-2\baselineskip}

\end{document}